# Semiclassical Model for Calculating Exciton and Polaron Pair Energetics at Interfaces

Michael J. Waters, Daniel Hashemi, and John Kieffer

**Abstract.**

Exciton and polaron pair dissociation is a key functional aspect of photovoltaic devices. To improve upon the current state of interfacial transport models, we augment the existing classical models of dielectric interfaces by incorporating results from ab initio calculations, allowing us to calculate exciton and polaron binding energies more accurately. We demonstrate the predictive capabilities of this new model using two interfaces: (i) the boron subphthalocyanine chloride (SubPc) and $C_{60}$ interface, which is an archetype for many organic photovoltaic devices; and (ii) pentacene and silicon (100), which represents a hybrid between organic and inorganic semiconductors. Our calculations predict that the insertion of molecular dipoles at interfaces can be used for improving polaron pair dissociation and that sharp transitions in dielectric permittivity can have a stronger effect on the polaron pair dissociation than even the electron-hole Coulomb interaction.

**Introduction.**

In photovoltaic devices, photons excite charge carriers that must be separated and collected to generate current. In all materials, the excited electron and the pseudo- or quasi-particle hole interact. In most cases, this interaction is dominated by the Coulomb component, which acts to bind the excited electron and hole into a charge neutral effective particle: the exciton. When an exciton becomes split over and interface where the electron and hole are mostly located on opposite sides of the interface yet still bound, a polaron pair or charge transfer exciton[1] is formed. The polaron pair binding energy is one of the largest limiting factors to the performance of organic and hybrid organic/inorganic photovoltaic devices. It subtracts from the open circuit voltage and directly limits the device current via the dissociation rate of polaron pairs at the accepter/donor interface (shown in Fig. 1).[2, 3]

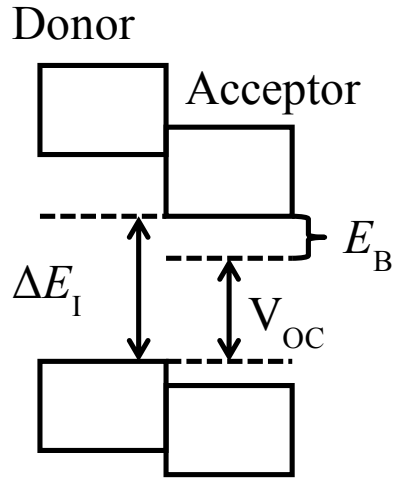

Figure 1. The band gaps and band alignment at a hypothetical interface. The $\Delta E_I$ is the energy for an electron to overcome when excited across the interface, based on the band edges alone. The polaron pair binding energy $E_B$ directly limits the electronically available $V_{OC}$.

Progress with respect to the theoretical description of polar pair dissociation kinetics has been slow, limiting the accuracy of device scale transport models. The most commonly used models are based the Onsager-Braun models.[2–4] Onsager's original model was developed for the electric field assisted dissociation of ions in solution,[5] which Braun applied to the dissociation of excitons and charge transfer states.[6] While some shortcomings of these models in describing polaron pair dissociation rate have already been pointed out in the literature,[7] we focus on creating a combined approach electronic structure and long-range classical electrostatics approach to improving polaron pair binding energy approximations as the first step in the creation of more accurate polaron pair dissociation and recombination rate models.

As a background to our work, the two prototypical exciton types (Frenkel and Wannier) are reviewed to better understand the effect excitonic differences have on polaron pair behavior. Following this, we review of the theory as its stands for excitons and polaron pairs at heterojunctions and present an overview of classical electrostatic interactions at dielectric interfaces. In the second part, we apply our methods to an interface between $C_{60}$ and boron subphthalocyanine chloride (SubPc) as an example of Frenkel-Frenkel polaron pair. Finally, we examine the interface between pentacene and a silicon (100) surface as an example of Wannier-Frenkel pair.

**Wannier Excitons.**

In traditional semiconductors made of inorganic materials, photo-excited electrons and their holes behave, for most intents and purposes, as separate particles. However, weakly bound exciton states, called Wannier or Wannier-Mott excitons, are characterized by hydrogenic states delocalized over many unit cells that travel via wave propagation. The energy levels are quantized as:

$$E_n = \frac{\mu e^4}{8(\epsilon h n)^2} = \frac{\mu^*}{\epsilon_r^2} \frac{1}{n^2} \frac{m_e e^4}{8(\epsilon_0 h)^2} = \frac{\mu^*}{\epsilon_r^2} \frac{1}{n^2} R_y \approx \frac{\mu^*}{\epsilon_r^2} \frac{1}{n^2} \times 13.605 \text{ eV}$$

*(Equ. 1)*

Where $\mu = \frac{m_e^* m_h^*}{m_e^* + m_h^*}$ and $\mu^* = \frac{1}{m_e}\left(\frac{m_e^* m_h^*}{m_e^* + m_h^*}\right)$ are the reduced effective mass and scaled reduced effective mass of the electron-hole pair. $n$ is the quantum number of the exciton. The Bohr radius for the ground state, which gives an approximation for the minimum exciton size is:

$$a_r = \frac{\epsilon h^2}{\mu \pi e^2} = \frac{\epsilon_r}{\mu^*} \frac{\epsilon_0 h^2}{m_e \pi e^2} = \frac{\epsilon_r}{\mu^*} a_0 \approx \frac{\epsilon_r}{\mu^*} \times 0.52918 \text{ Å}$$

*(Equ. 2)*

For most classical semiconductors, which have small effective masses and large dielectric constants, the binding energies are small (usually less than $k_B T$) and the radii are large (a few nm). For example, the above formulas applied to silicon yield a binding energy of 55 meV and a radius of 21 Å.

**Frenkel Excitons.**

In the more recently popularized organic semiconductors, excited electrons are by comparison strongly bound to their holes, forming Frenkel excitons. This binding leads to the exciton pseudo-particle where the electron and hole travel together as an effective neutral particle. The strong binding and localization mean that Frenkel excitons tend to travel via tunneling from site to site. The binding interaction is usually described as purely Coulomb, as with the Wannier type. The difference is that the strongly localized electron and holes require treatment of the actual wavefunction rather than treating them as perturbations of valence and conduction bands. A good approximation of the binding energy is:

$$E_{bind} = \frac{e^2}{4\pi\epsilon} \iint \frac{|\psi_h(\vec{r}_h)|^2 |\psi_e(\vec{r}_e)|^2}{|\vec{r}_h - \vec{r}_e|} d\vec{r}_h d\vec{r}_e$$

*(Equ. 3)*

which uses the dielectric permittivity of the bulk material to account for the screening of other atoms and molecules. The exciton radius is oftentimes computed based on a measured binding energy according to

$$r_{e-h} = \frac{e^2}{4\pi\epsilon E_{bind}}$$

*(Equ. 4)*

However, the exact spatial distribution of charges is disregarded, which can lead to error. As a case in point, consider that two spherically symmetric co-centered Gaussian charge distributions have a finite binding energy but no distance between them. (See Appendix A for derivation.)

In the case of the Wannier exciton spatial confinement of the wave function comes from the Coulomb interaction between electron and hole, whereas in the case of Frenkel excitons, it comes in part, from the spatial extent of the molecular orbitals. If we consider a polaron pair to be an exciton split over an interface, then three possible pairings of exciton type can be made: Frenkel-Frenkel, Wannier-Wannier, and Frenkel-Wannier.

In the case of the Frenkel-Frenkel pairs, most authors treat the exciton binding energy using the Mulliken rule where the lowest energy optical absorption peak is attributed to the charge transfer exciton[8–10] For large donor-acceptor distances, $R$, the Mulliken rule is:

$$hv = IP(D) - EA(A) - \frac{q^2}{4\pi\epsilon R}$$

*(Equ. 5)*

where $hv$, $IP(D)$, and $EA(A)$ are the photon energy, donor ionization potential, and acceptor, electron affinity, respectively. The final term accounts the for the Coulomb binding between the electron and hole. Although some have included classical image potentials from dielectric interfaces into the binding energy considerations,[11] for the most part, it seems that exciton binding energies are calculated solely with the Coulomb interaction between the electron and hole states. These are then fed into the Onsager-Braun model to calculate dissociation rates. However, a Poole-Frenkel model may be more accurate.[4, 12, 13]

In the case of Wannier-Wannier polaron pairs, there has been much analytical work on solving exciton Hamiltonians in the presence of dielectric interfaces with image potentials near interfaces.[14–18] Unfortunately, in these calculations, the excitons have been constrained to reside in one material by applying an infinite potential barrier rather than allowing the excitons to dissociate across the interface to form polaron pair states. It would be preferable that finite potential barriers, which better relate to the differing band structures of the two materials, spatially separate the carriers.

In the case of a hybrid interface consisting of Wannier material on one side of the interface and a Frenkel material on the other side, neither model can accurately describe the system. One may consider hydrogen like states trapped in a half space for one of the carriers, but then each position of the Frenkel-like carrier has a different set of quantized exciton binding energies. To further complicate the matter, the electric field dependence of the dissociation rate requires a full quantum mechanical treatment. Considering the small size of Frenkel excitons relative to Wannier excitons, a hybrid polaron pair might be treatable as a hydrogenic atom in two adjacent dielectric half spaces where one half space has a finite potential. Some work has been done for hydrogen in a single dielectric material with an infinite half space which with some theoretical extension, may be applicable to hybrid interfaces.[19, 20]

**Classical Electrostatics Effects.**

We examine the idealized case of carrier interactions near a planar interface, because most photovoltaic devices tend to be planar, as they are fabricated via spin coating or some sort of chemical or vapor deposition method. Since this is a straightforward electrostatics problem, some solutions already exist.[21] The following relationships are derived from these known solutions in the context of device performance in the form of energetic effects on polaron pair binding energy and are summarized here, for a more complete description, see Appendices B-E.

In the simplest configuration, a single carrier (of charge $q_1$), rests in material 1 near the interface with material 2. This charge creates a potential field, $\phi^1$, described by the piecewise function:

$$\phi^1(\vec{r}) = \begin{cases} \phi_1^1(\vec{r}), z > 0 \\ \phi_2^1(\vec{r}), z < 0 \end{cases}$$

*(Equ. 6)*

$$\phi_1^1(\vec{r}) = \frac{q_1}{4\pi\epsilon_1}\left[\frac{1}{|\vec{r}-\vec{r}_1|} + \frac{\alpha_1}{|\vec{r}-\vec{r}_1{}'|}\right]$$

*(Equ. 7)*

$$\phi_2^1(\vec{r}) = \frac{q_1}{4\pi\epsilon_2}\left[\frac{\beta_1}{|\vec{r}-\vec{r}_1|}\right]$$

$$\alpha_1 = \frac{\epsilon_1-\epsilon_2}{\epsilon_1+\epsilon_2}, \quad \beta_1 = \frac{2\epsilon_2}{\epsilon_1+\epsilon_2}$$

*(Equ. 8)*

where interface is a $z = 0$ at and material 1 and 2 are respectively in the positive and negative $z$ directions. $\vec{r}_1$ is the position of charge 1 and $\vec{r}_1{'}$ is the position mirrored across the interface, *i.e.* $(x, y, z) \to (x, y, -z)$. The dielectric permativities of materials 1 and 2 are $\epsilon_1$ and $\epsilon_2$, repectively. The second term in the brackets containing the mirror position represents the potential field due the areal bound charge density induced by the original charge as seen in Figure 2.

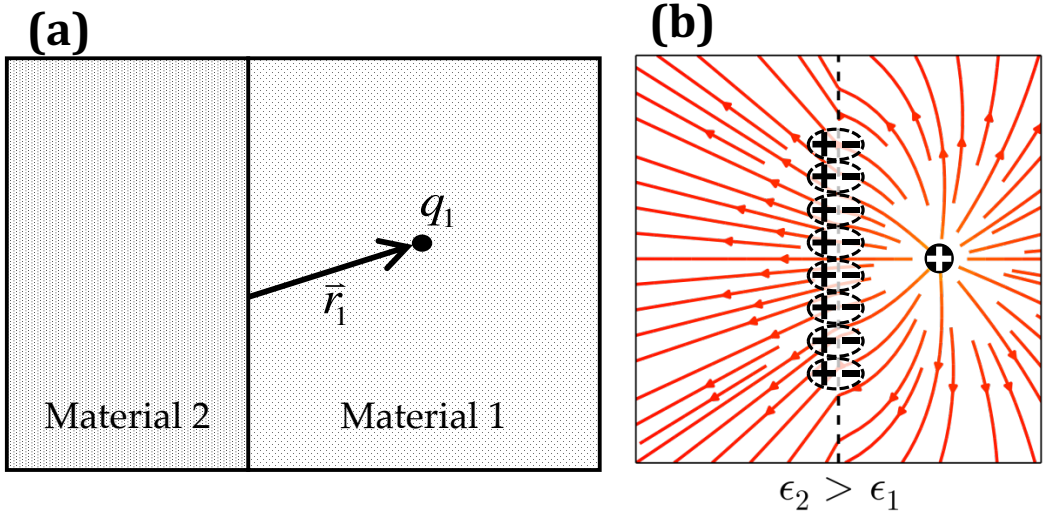

Figure 2. (a) A charge of $q_1$ in located in material 1 (b) The electric field lines extending from a positive charge in material 1. With the permittivity of material 1 less than material 2, bound negative charges are induced a the interface.

The electric field of the areal bound charge density, represented in Equation 9 below, acts on the carrier to draw it into the material with the higher dielectric constant. The charge interacts effectively with its own image charge.[11]

$$E_1^B(\vec{r}) = \frac{q_1}{4\pi\epsilon_1}\left[\alpha_1 \frac{\vec{r}-\vec{r}_1{'}}{|\vec{r}-\vec{r}_1{'}|^3}\right]$$

*(Equ. 9)*

By integrating the electric field due to the bound charges, the self-polarization potential due to the interface is obtained.[22] the causes the carrier to move away from the interface and deeply into material 1, and is given by Equation 10, where $h_o$ is the initial distance to the interface. It should be noted that this potential is proportional to the difference in the dielectric constants and can be either positive or negative.

$$\int_{\infty}^{h_o} -q_1 E_1^B \cdot \hat{z} dh = U_{self} = \frac{q_1^2}{16\pi\epsilon_1 h_o}\left(\frac{\epsilon_1 - \epsilon_2}{\epsilon_1 + \epsilon_2}\right)$$

*(Equ. 10)*

In the case of a polaron pair, two carriers are on either side of the interface between material 1 and material 2 as in Figure 3. The electrostatic potential between the two carriers is derived by evaluating the potential at a second charge, $q_2$, in material 2. The surprisingly simple result is described by Equation 11, where $d$ is the distance between the charges.

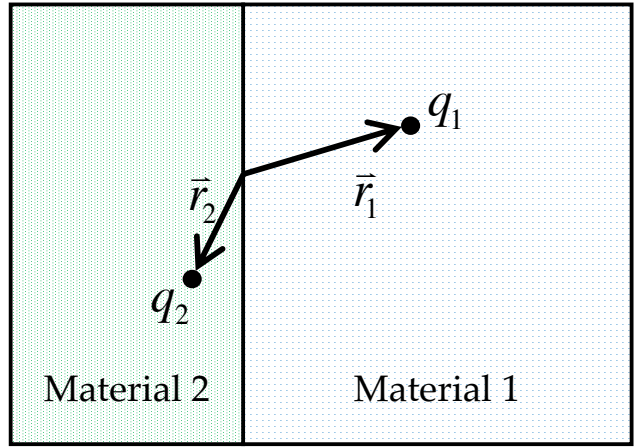

Figure 3. Two charges, $q_1$ and $q_2$, are located on either side of the interface.

$$U_{charge-charge} = \frac{q_1 q_2}{4\pi d (\epsilon_1 + \epsilon_2)/2}$$

*(Equ. 11)*

Interestingly, this is result is valid no matter what the positions of the two charges are, so long as they are on different sides of the interface. This result can be easily extended to diffuse charges on either side of the interface in Equation 12,

$$U_{charge-charge} = \frac{1}{4\pi(\epsilon_1+\epsilon_2)/2} \int_{z>0}^{z\to\infty} d^3r_1 \int_{z<0}^{z\to-\infty} d^3r_2 \left[\frac{\rho_1(\vec{r}_1)\rho_2(\vec{r}_2)}{|\vec{r}_2-\vec{r}_1|}\right]$$

*(Equ. 12)*

where $\rho_1(\vec{r}_1)$ and $\rho_2(\vec{r}_2)$ are the charge densities on either side of the interface.

In some cases, there may be a local polarization of the material near an interface as a result of epitaxy. In the case of perovskites, a thin layer of polarized material can develop as result of epitaxial strain near the interface, while being relieved in the bulk due to misfit dislocations. In the case of an amorphous SubPc film deposited on $C_{60}$, the preferred molecular orientation of the polar SubPc molecules can lead to local polarization.[23] For a thin polarized layer between the two materials the resulting potential field exhibits a discontinuity at the interface. The corresponding change in potential is:

$$\Delta\phi_{2\to 1} = \frac{\sigma_z}{2}\left[\frac{1}{\epsilon_1}+\frac{1}{\epsilon_2}\right]$$

*(Equ. 13)*

Where $\sigma_z$ is the areal dipole moment density perpendicular to the interface. For a full derivation, see Appendix E. Depending on the configuration of the system of interest, this may aid in the dissociation of excitons to form polaron pairs or reduce the polaron pair binding energy depending on the relative location of the dipoles to the interface. In perovskites, a more complex situation may occur where flexoelectric polarization density decays with distance from the interface as misfit dislocations relieve a lattice mismatch.[24–26] The net result is a small built-in field that can increase or decrease the polaron pair binding energy.

**Methods.**

For the SubPc/$C_{60}$ interface, previous work has shown the preferred orientation of SubPc on $C_{60}$ (111) surfaces is the ball-in-cup configuration. In this configuration, it is ultimately favorable for an electron to be adiabatically excited from the HOMO of the SubPc to the LUMO of the $C_{60}$.[27] This is also observed with phthalocyanine and $C_{60}$.[28] However, it was not known if room temperature thermal motion would significantly affect the polaron pair binding energy. To investigate this, the thermal motion of an isolated SubPc sitting on the (111) surface of $C_{60}$ is simulated using *ab initio* MD. The $C_{60}$ (111) surface comprises one unit cell containing four $C_{60}$ molecules in a single layer as seen in Figure 4. Before being used as input for *ab initio* calculations. isolated molecular structures are created in Avogadro and relaxed using the built-in potentials.[29]

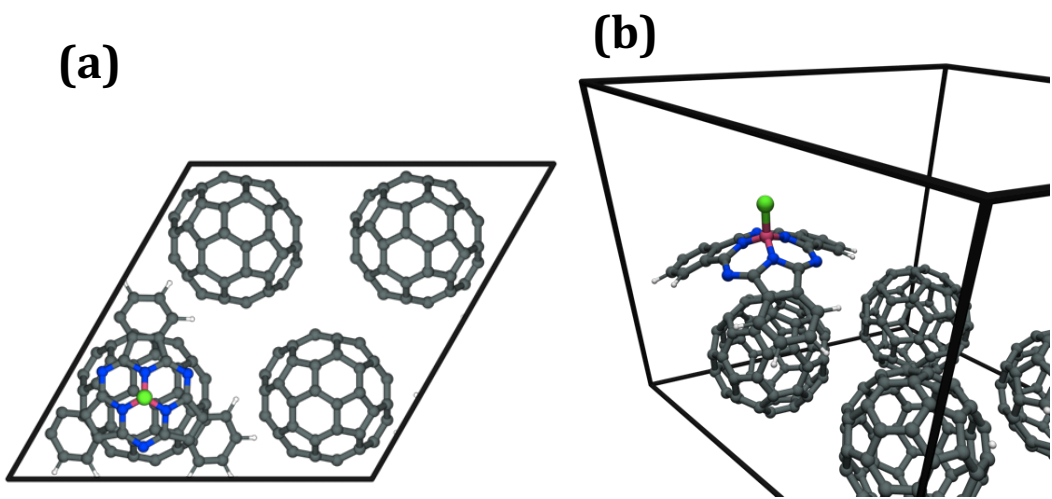

Figure 4. (a) Top down view of the SubPc molecule and $C_{60}$ surface. (b) perspective view of the same SubPc molecule and $C_{60}$ surface.

*Ab initio* molecular dynamics simulations are done with VASP (version 5.3.3), using the PAW method[30] with the PBE exchange-correlation functional.[31] Augmented plane waves with a cutoff energy of 120 eV form the basis set. Van der Waals interactions are accounted for using the VdW-DFT approach developed in Refs. 32 and 33 without PBE correlation correction. In the initial configuration, a 7 Å vacuum gap is added to the super cell between the chlorine atom of the Subpc molecule and the bottom of the $C_{60}$ to minimize interactions with the periodic images in the surface normal direction (z-direction). Nuclear motion is integrated with a time step of 1 fs. A Nosé-Hoover thermostat is employed with a target temperature of 300 K and an equilibration time of ~ 3 ps. Once the system is equilibrated, simulations are continued for another 12 ps to capture the relative position space of the $C_{60}$/SubPc pair. The relative orientation and position of the SubPc molecule at each time step are calculated using Procrustes (also known as Kabsch) analysis, which is based on computing the least squares rotation matrix after the centroid motion has been accounted for.[34]

A series of single-point electronic structure calculations are performed for the SubPc molecule and only the $C_{60}$ molecule immediately beneath it as function of centroid separation distance between the two. For these single point calculations, Gaussian 09, Revision C.01 is used with the B3LYP hybrid functional and the 6-31G(d) basis set. For each centroid distance, a Coulomb integral is evaluated on the density mesh automatically generated by Gaussian. To simplify the intensive six-dimensional integral, only cells inside the isodensity surface containing 99% of the HOMO or LUMO states are used. Effective cell densities are corrected for the loss of the remaining 1%. While this does not change the asymptotic computational complexity of the calculation, it results in a 222-fold speedup at the cost of less than 5 meV error. The Coulomb contribution to the polaron pair binding energy at each time step is interpolated among the centroid distance series of single point calculations using a cubic spline.

Two (001) silicon slabs were created, one with a bare surface and the other one with hydrogenated surface, both consisting of 27 layers of silicon atoms (~35 Å thick) in a 2 x 2 unit cell slab as shown in Figure 5. This large number of atoms is required to minimize vertical quantum confinement effects on the electronic structure, which typically is inversely proportional to the square of the slab thickness. A vacuum gap of 10 Å between the highest and lowest atoms is included to remove periodic interactions in the surface normal direction.

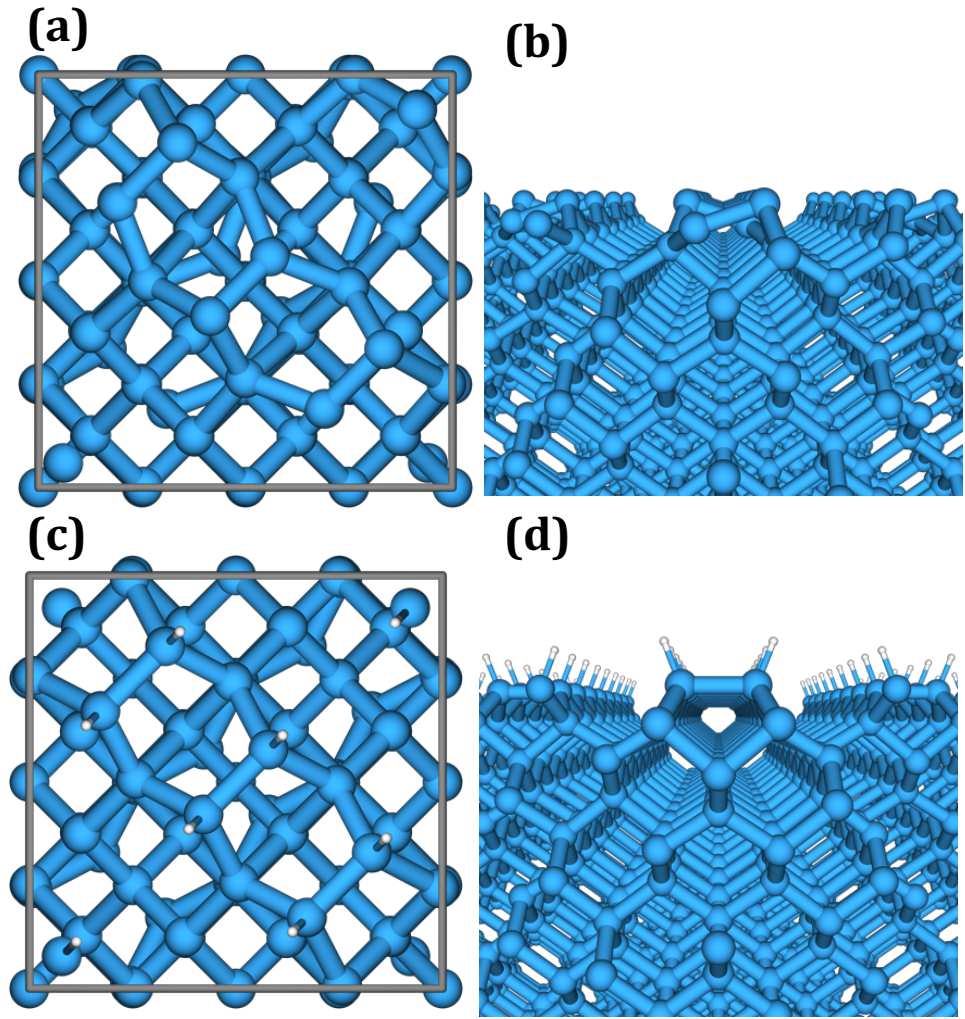

Figure 5. (a) The clean silicon (001) surface is viewed top down and (b) along the alternating dimer ridges. (c) The hydrogenated surface is also viewed top down and (d) along the hydrogen terminated dimer ridges.

All electronic structure calculations of the slabs are carried out using VASP version 5.3.5. The augmented plane wave basis set is cut off at 400 eV and electronic relaxations are converged to an energy difference of 10 μeV. *k*-point grids are automatically generated using the Monkhorst-Pack scheme. The atoms of both slabs are relaxed using the PBE exchange-correlation density functional until individual atomic forces no longer exceed 10 meV/Å.[31] The more costly, split-range hybrid method HSE06 is used for more accurate electronic structure calculations.[35]

Total energy calculations of bulk silicon are found to converge within 150 μeV/atom using an 8 x 8 x 8 *k*-point grid and the aforementioned plane wave cut-off. The lattice parameter of silicon calculated using HSE06 is used for the construction of silicon slabs. The electronic structure of bulk silicon is calculated for comparison to the projected band structures of the 2 x 2 slabs using the conventional cell with a 16 x 16 x 16 *k*-point grid. The projected band structures for the 2 x 2 slabs are computed using a 4 x 4 x 1 *k*-point grid since the slab is not periodic in the surface normal direction.

For the final structures with the pentacene molecule, the 2 x 2 unit cells slabs are duplicated twice in each surface direction to provide enough space in the lateral directions for the pentacene molecule. The molecule is placed in line with the dimer ridges and relaxed while Van der Waals interactions between the surface and the pentacene molecule are accounted for using the VdW-DFT method developed in Refs. 32 and 33 with removed PBE correlation correction. Only one *k*-point and molecule are used since each structure is computationally costly containing over 900 atoms. Wave functions are extracted using the WaveTrans[36] code developed for Ref. 37. Coulomb integrals are calculated as they were for the SubPc/$C_{60}$ in the previous section. Dielectric constants of 3.61 for pentacene[38] and 11.7 for silicon[39] are used.

**Results and Discussion**

From tracking the relative position and orientation of the SubPc molecule, it is found that the SubPc molecule strays less than 10° from surface normal with a mean of only 3.6°, only the centroid distance to the nearest $C_{60}$ molecule is considered for Coulomb integrals. The distribution of the centroid distributions is shown in Figure 6.

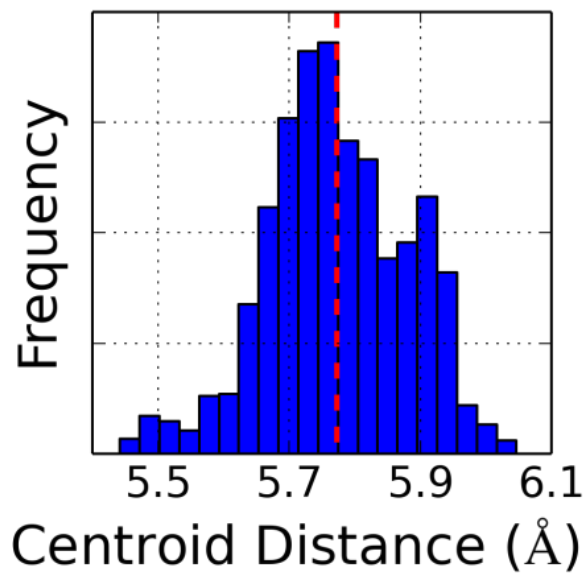

Figure 6. The distribution of the centroid distances of the SubPc molecule and nearest $C_{60}$ molecule with the mean centroid distance shown as a dotted red line.

For each single-point calculation, Coulomb integrals are evaluated to obtain the Coulomb contribution to the polaron pair binding energy as a function of the centroid distance, as shown in Figure 7. The Coulomb contribution to the polaron pair binding energy at each time step was interpolated from the series single-point calculations to get a mean and standard deviation of the Coulomb contribution to the polaron pair binding energy.

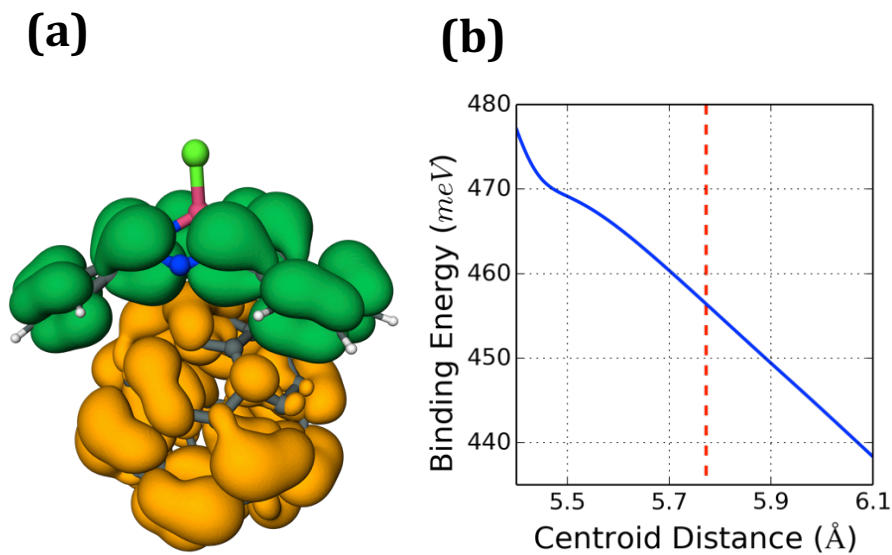

Figure. 7. (a) The HOMO of the SubPc is shown green and the LUMO of the $C_{60}$ is shown gold. Both are shown as isodensity surfaces containing 80% of the state. (b) The Coulomb contribution to the polaron pair binding energy is plotted as function of the centroid distance between the SubPc and $C_{60}$ molecules. The mean centroid distance shown as a dotted red line.

The most important result of the *ab initio* MD simulation is the small effect that thermal motion has on the Coulomb contribution to the polaron pair binding energy. In fact, the difference between the Coulomb energy associated with the equilibrium position and that of any position within the thermal distribution is safely smaller than $k_B T$. The insignificance of thermal motion in this case is further underscored by the lack of neighboring molecules to hinder motion of the SubPc molecule in our model surface. The charge center approximation often used by simpler models does well to first order, but in the current race for highly efficient devices it is not likely accurate enough.

Table 1. The Coulomb contribution to the polaron pair binding energy from various calculation methods.

| | |
|---|---|
| Charge center (Equation. 3) | 537 meV |
| Coulomb integral at equilibrium (Equation. 4) | 457 meV |
| Coulomb integrals over thermal distribution | 456 ± 6 meV |

The Coulomb integral results can be combined with an estimate on the upper bound of the effect polarized interface (the net ordering of dipoles at the interface) and the self-polarization energies to gain a better understanding of their relative contributions. The Procrustes analysis shows that the SubPc molecules do not significantly tilt from vertical, which means that their dipole moments are essentially normal to the interface.

We assume the upper bound interface dipole density where half of the $C_{60}$ sites have a vertical SubPc molecule. The upper bound polarized interface energy is large enough to non-trivially weaken the polaron pair binding energy. In this case the polarized interface reduces the polaron pair binding energy. The net effect of the interfacial termination depends on the dipole orientation. Hence, in another system, dipoles at the interface can enhance this bonding energy. In conjunction with the work in Ref. 40, the effect of the molecular dipole moment on the I-V curve of a real SubPc/$C_{60}$ device is modeled for five cases: no molecular alignment, ¼ of $C_{60}$ sites covered with SubPc molecules, ½ of $C_{60}$ sites covered with SubPc molecules and dipole inversions the last two cases. The resulting simulated IV curves are shown in Figure 8, it can be seen that alignment of the molecular dipoles can have a large effect on device performance.

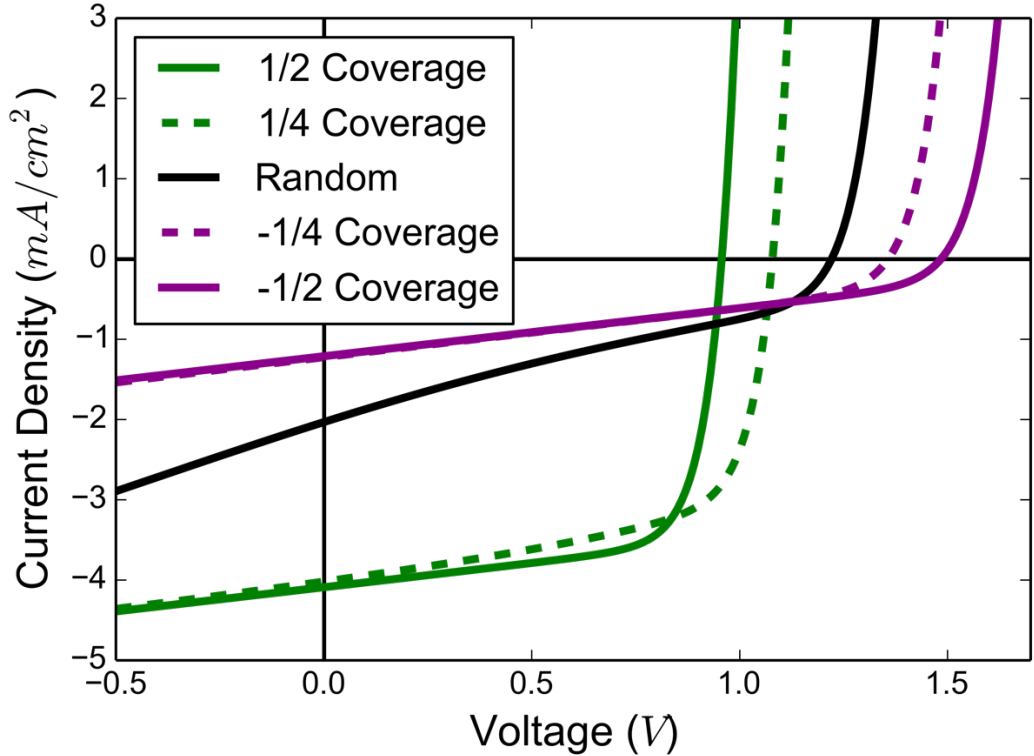

Figure 8. The modeled I-V curves for a SubPc/$C_{60}$ device with half of the $C_{60}$ sites covered with SubPc molecules (solid green line), one quarter of the $C_{60}$ sites covered with SubPc molecules (dashed green line), random SubPc molecule orientation (solid black line), one quarter of $C_{60}$ sites covered with inverted SubPc molecules (dashed purple line), and half of $C_{60}$ sites covered with inverted SubPc molecules (solid purple line).

For the self-polarization energy, we estimate the interface to be halfway between the closest hydrogen and carbon atoms of the SubPc molecule and the adjacent $C_{60}$ molecule, respectively. The effects self-polarization on the polaron pair binding energy are collected in Table 2 below. Unsurprisingly, the small difference in permittivity between SubPc and $C_{60}$ gives a negligible self-polarization energy contribution.

Table 2. All of the semi-classical corrections to the polaron pair binding energy for the SubPc/$C_{60}$ interface are tabulated. *The polarized interface energy is an upper bound.

| | |
|---|---|
| Self-Polarization Energy (Equ. 1) | 17 meV |
| Polarized Interface Energy* (Equ. 4) | -237 meV |
| Coulomb Contribution (Equ. 3) | 456 meV |
| Total Polaron Pair Binding Energy | 236 meV |

To demonstrate the effect of the dielectric constant, the same device as in Figure 8 is considered, except that the dielectric permittivity of the $C_{60}$ is varied. The dielectric constant of $C_{60}$ is arbitrarily increased from the experimental value of 5.0 to 15.0 in Figure 9 below. As the Coulomb interaction weakens, the self-polarization energy grows holding holes in the SubPc more strongly to the interface. The self-polarization energy grows more slowly than the Coulomb energy decreases due to the smaller leading coefficient in Equation. 10. The net result is that increasing the dielectric permittivity improves device performance but there is no significant improvement beyond a relative permittivity of 10.0.

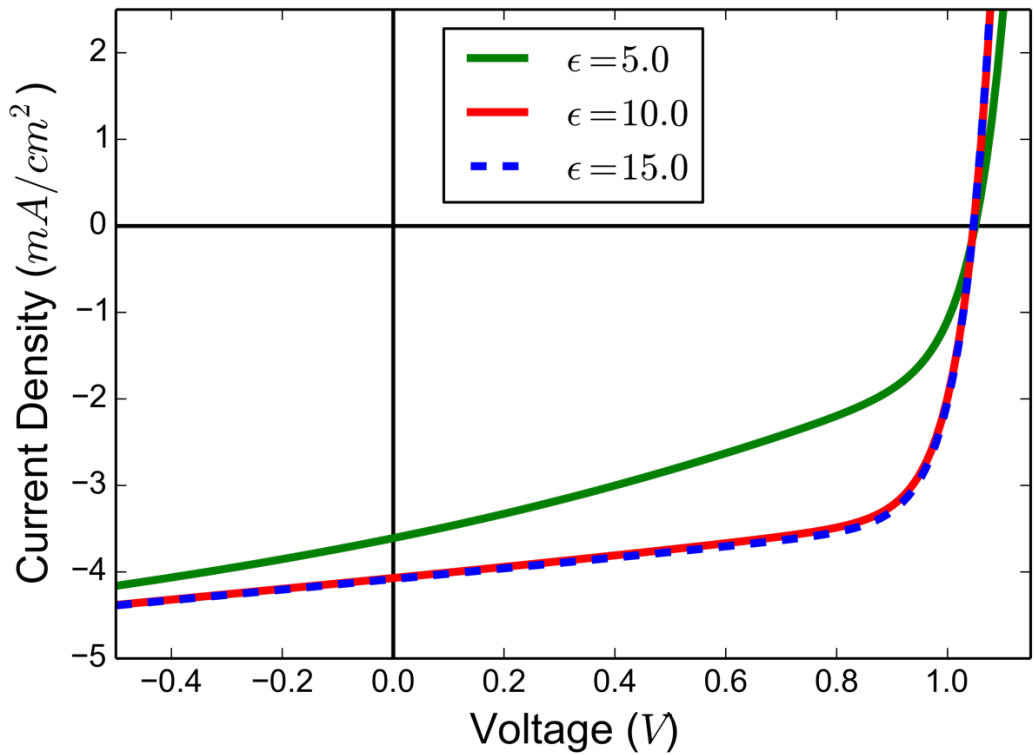

Figure 9. Simulated I-V curves of a hypothetical SubPc/$C_{60}$ devicewhere the relative permittivity of $C_{60}$ is the experimental value (solid green line), 10.0 (solid red line), and 15.0 (dashed blue line).

For bulk silicon, the lattice parameter and bulk modulus are calculated using both HSE06 and PBE. Both agree well with experimental data for structure and mechanical properties. (Table 3) As usual, PBE (and DFT in general) underpredicts the band gap, which is why HSE06 was used for electronic structure calculations despite the significant increase in computational cost.

Table 3. The calculated properties of silicon compared with experimental values.

|  | PBE | HSE06 | Experimental |
|---|---|---|---|
| Lattice Parameter (Å) | 5.4685 [+0.72 %] | 5.4332 [+0.07 %] | 5.4293[41] [0K] |
| Bulk Modulus (GPa) | 92.5 [-6.28 %] | 99.8 [+1.11 %] | 98.7 [233 K] Calculated from Ref. 42. |
| Band gap (eV) | 0.579 [-50.44 %] | 1.158 [-1.03 %] | 1.170[43] [0K] |

The relaxed bare surface exhibits the p(2 x 2) buckled dimer reconstruction and the hydrogenated surface the symmetric dimer reconstruction. There are several reconstructions of the silicon (001) surface with different buckling orders of the dimers but energy differs between these less than $k_B T$ per dimer.[41] The projected band structures of the two surfaces reveal distinctly different electronic structures near the band gap. The clean surface has two distinctive bands that sit in the middle of the band gap as seen in Figure 10a. These bands can clearly be seen to be surface states associated with the surface reconstruction in Figure 10c and Figure 10d. The band gap for the clean surface is 0.615 eV (1.268 eV if the surface states are ignored) and 1.209 eV for the hydrogenated surface. These band gaps are slightly larger than the bulk band gap, which is easily attributable to quantum confinement in the z-direction.

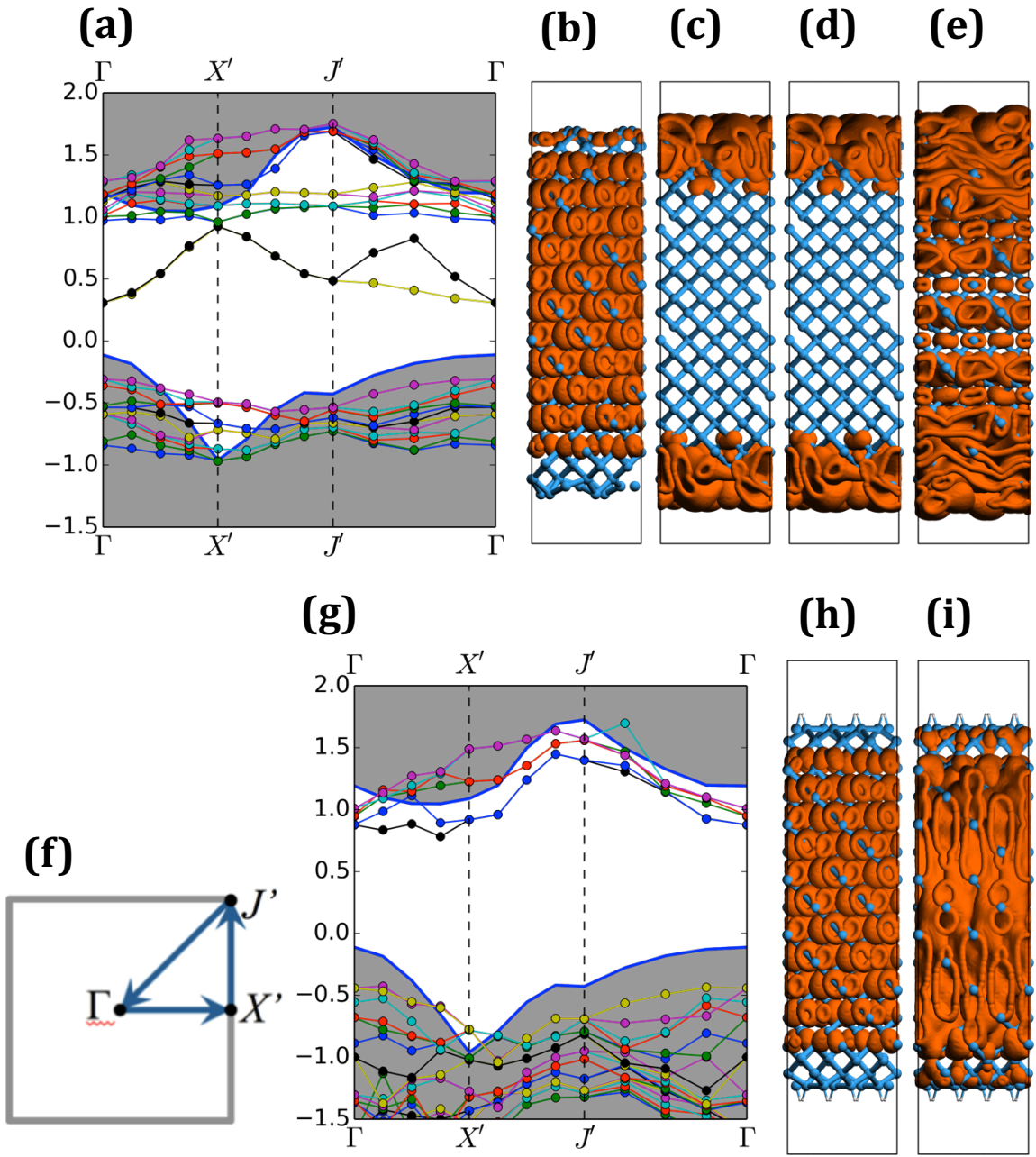

Figure 10. The projected band diagrams of the clean surface and the hydrogenated surface are shown in (a) and (g) respectively. At the Γ-point, the valence band, the first surface state, the second surface state, and the conduction band for the clean surface are shown in (b), (c), (d), and (e), respectively. Also At the Γ-point, the valence band and the conduction band for the hydrogenated surface are shown in (h) and (i), respectively. (f) shows the band diagram path through the Brilliouin zone use in (a) and (g).

For the pentacene functionalized surfaces, we found that the pentacene molecule adsorbs to the clean surface in the A-1 sub-type single symmetric dimer of Choudhary *et al.*[42] Meanwhile on the hydrogenated surface, the molecule remains flat and aligned with the dimer ridge. Both surfaces can be seen in Figure 11.

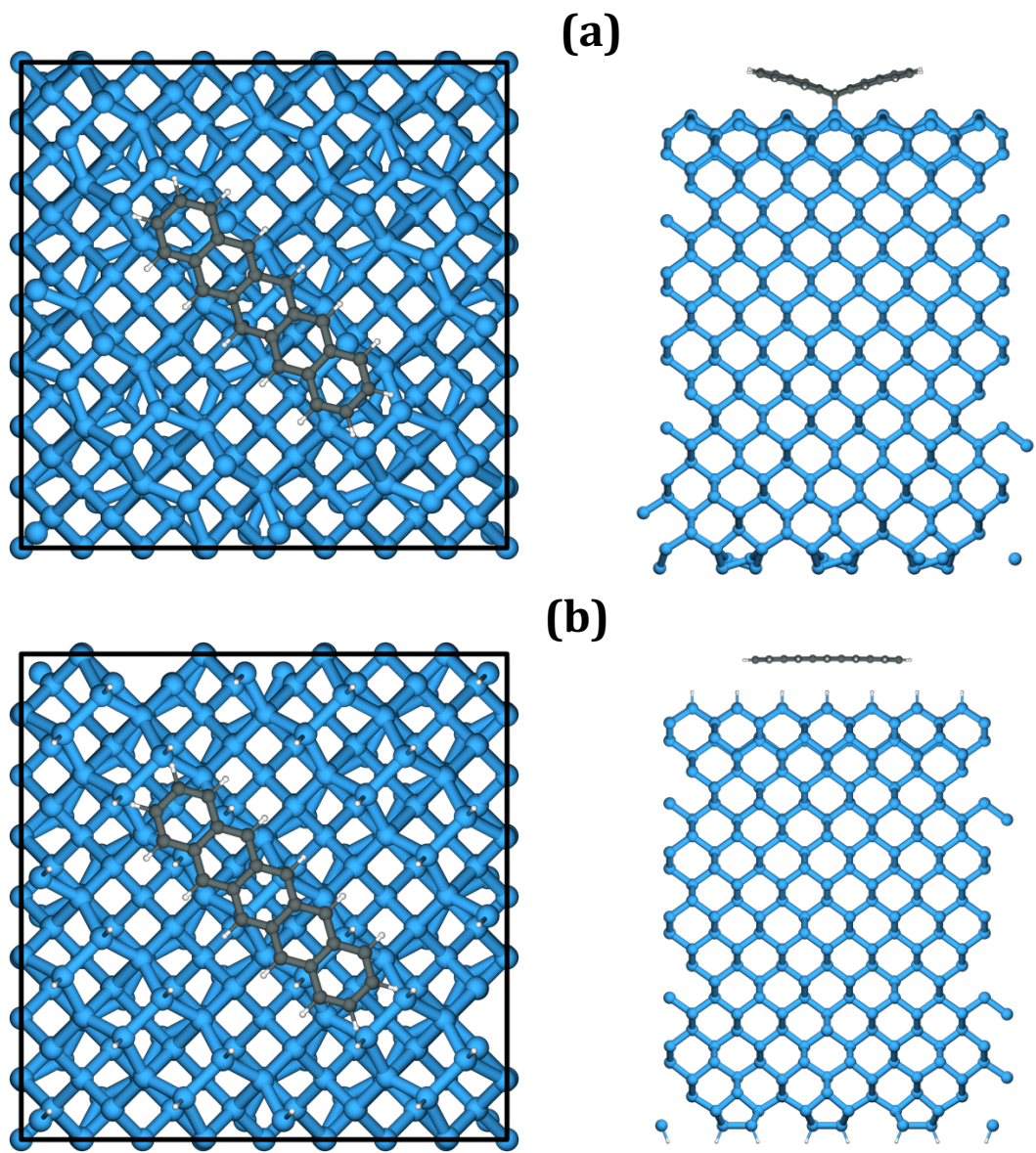

Figure 11. Pentacene adsorbed to a pristine (a) and a hydrogenated (b) Si (111) surface. In the latter case, the pentacene remains unreacted.

The electronic structures of the two surfaces are very different. The hydrogenized surface displays essentially independent electronic states between the silicon slab and pentacene molecule. The highest filled state (HFS) is simply the HOMO for pentacene molecule essentially unperturbed from the isolated molecule. The second highest filled state (HFS-1) is the valence band of the silicon slab. The lowest unoccupied state (LUS) is the conduction band of silicon. The HFS/LUS gap was found to be 1.256 eV. The state corresponding to the LUMO of the pentacene molecule is 478 meV above the LUS. These states are shown below with their corresponding DOS in Figure 12.

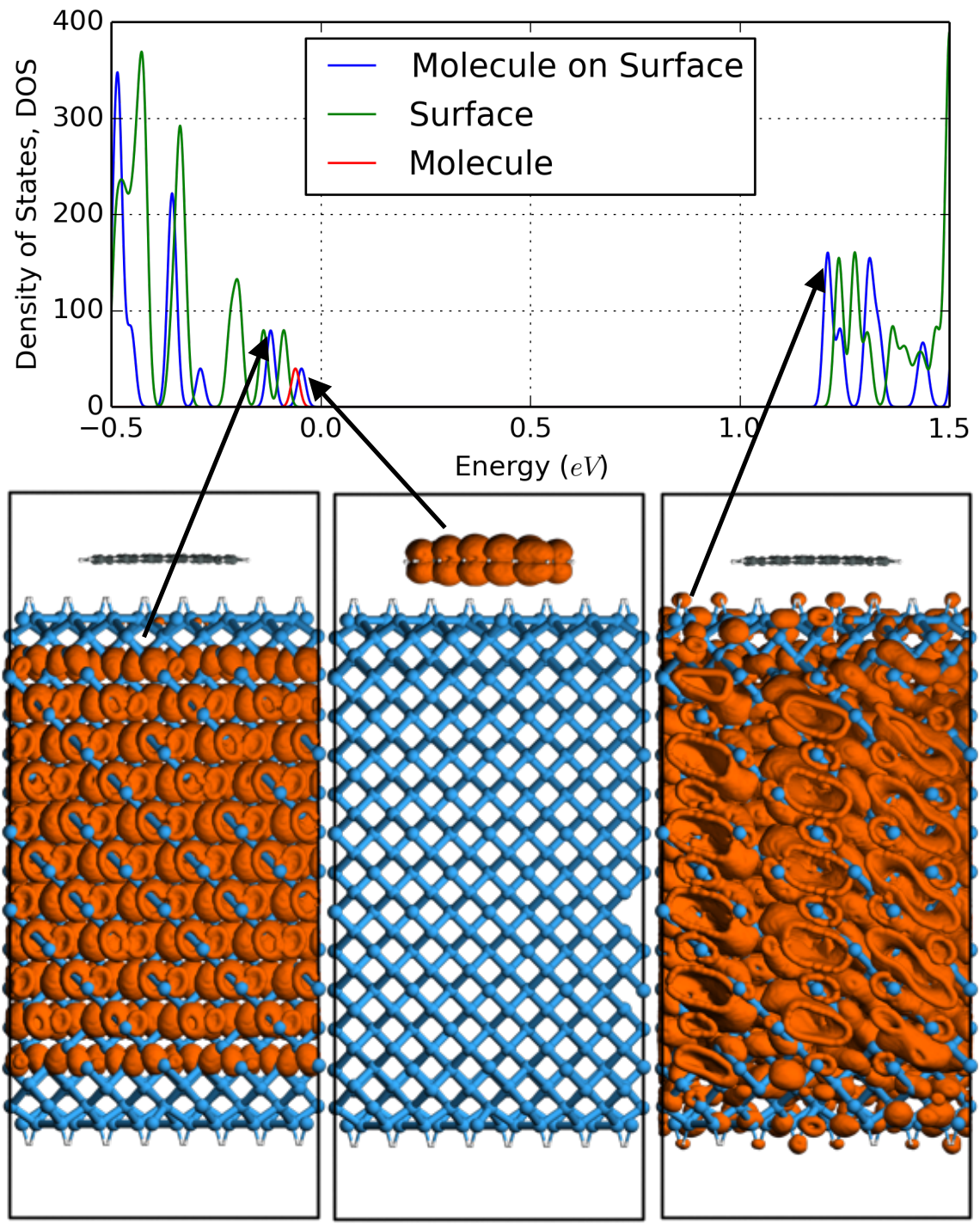

Figure 12. On top, the DOS of the hydrogenated surface with the pentacene molecule. From the left to right on the bottom, the HFS-1, HFS, and LUS which for an isolated surface correspond to the valence band of silicon, the HOMO of pentacene and the conduction band of silicon, respectively.

For the molecule on the clean surface, the reaction between the two completely changes the nature of the silicon surface states and the MOs of the pentacene molecule. The HFS now resides on the distorted pentacene molecule and penetrates into the surface. In fact, HFS-6 is the highest state with obvious distortion in relation to the pentacene molecule. These can be seen in the DOS shown in Figure 13. In the other direction, the LUS is the relatively undisturbed surface state on the opposing side of the silicon slab. The LUS+1 has a similar mixing of surface state and MO as the HFS. The lower unoccupied states all reside at the surfaces. In fact, the lowest unoccupied state residing mostly in the bulk of slab is LUS+5. The HFS/LUS gap is 0.668 eV and the HFS/LUS+1 gap is 0.606 eV. The HFS/LUS+1 gap better represents the interface transitions since both states actually reside there.

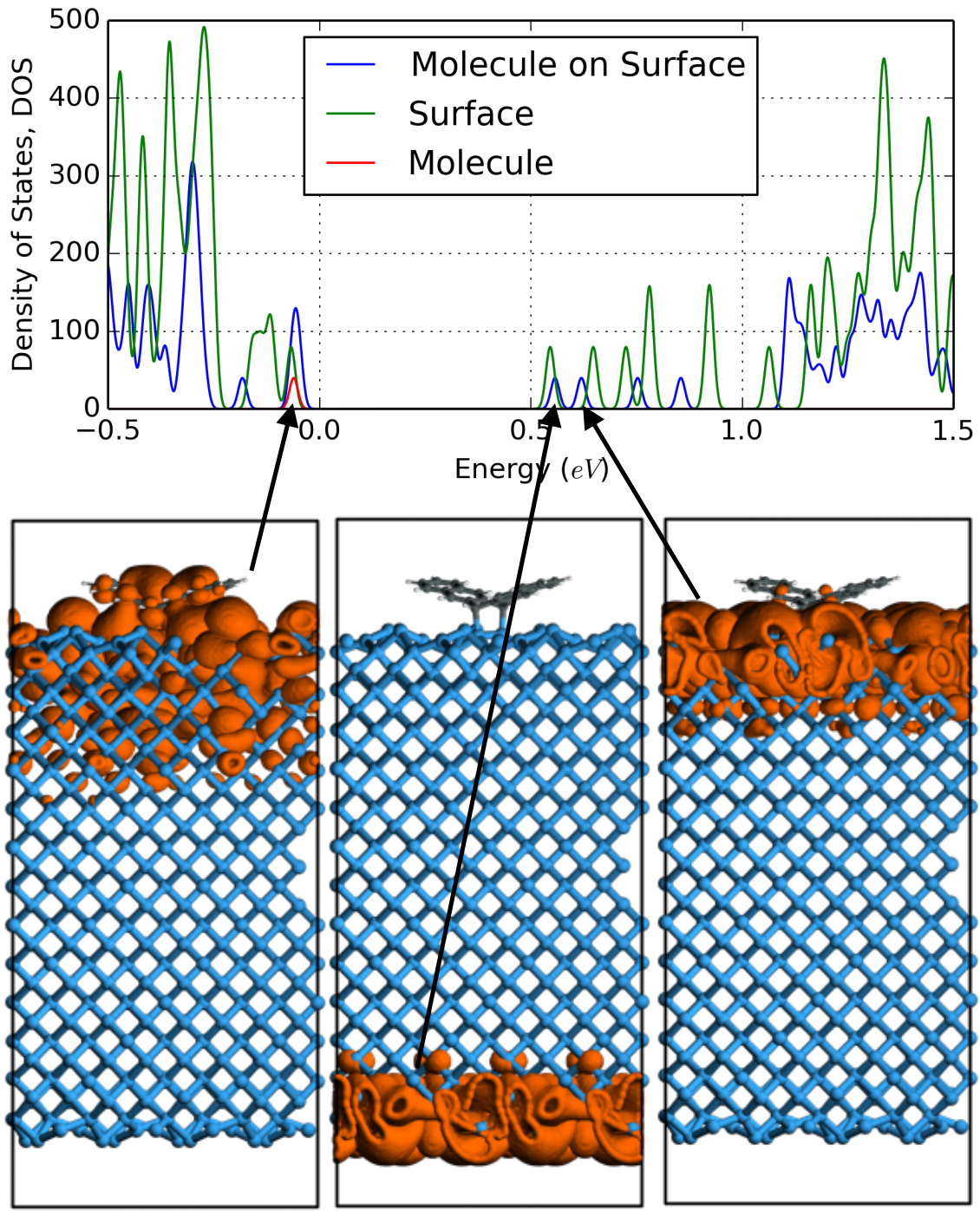

Figure 13. On top, the silicon surface with and without the molecule have very different DOS due to the states formed by the reaction of the molecule with the surface. From left to right on the bottom, the HFS, LUS, and LUS+1 are shown.

The calculation of the polaron pair binding energies using the semi-classical formulas presented in the classical electrostatics section require some approximation. A defined interface position is required for the application of self-polarization potential formula (Equation 10) because it relies on distance between the carrier and interface. The mean height of the hydrogen atoms at the interface is chosen for this because it also corresponds very closely to the height where the valence electron density decreases to half the bulk value. Any part of the states of interest that extended beyond this height is truncated because the carrier-interface potential is ill-defined for crossing the interface.[39] The truncation of the HFS and LUS across the interface was less than 0.005% of the total states. For the unhydrogenated interface, the centroids of the HFS and LUS+1 are both well inside the silicon and significantly distributed across the interface, which makes the self-polarization completely inapplicable. The Coulomb binding energy (Equation 12), while calculable, it is not as applicable when the states are mixing across the interface. Nonetheless, we present the Coulomb binding energy and self-polarization contributions for clean interface and the more applicable hydrogenated interface in Table 4. The significance of the self-polarization contribution is immediately evident for the hydrogenated interface, as it is more than double the Coulomb contribution. While previous models of hybrid interfaces address the Coulomb contribution reasonably, we demonstrate that the self-polarization contribution to polaron pair binding energy cannot be ignored.

Table 4. The contributions to the total polaron pair binding energy are given for the clean and hydrogenated surfaces with the estimate from the Renshaw model.

| Energies (meV) | Clean | Hydrogenated |
|---:|---:|---:|
| Self-Polarization Energy | N/A | 194 |
| Coulomb Integral | 201 | 87 |
| Total polaron pair binding energy | 201* | 281 |
| Renshaw Model[3] | 53 | 53 |

The electronic structure calculations used to obtain the HFS and LUS are self-consistent with respect the ground state electronic structure and do not take into account the quasiparticle interactions of the excited state. Taking these into account for the hydrogenated interface would likely result in the excited electron wave function in the silicon being shifted towards the hole on the pentacene molecule which would increase the Coulomb contribution somewhat. However performing quasiparticle calculation such as the GW on a system of ~ 1000 atoms is currently too computationally costly especially when our interest is in only a few states near the HFS. Furthermore, the purpose using less computationally costly electronic structure calculation with classical long-range electrostatic formulas is to avoid the computationally prohibitive methods while still arriving at good approximations.

**Conclusions.**

We attempted a more rigorous calculation of the electrostatic effects of interfaces on carriers, excitons, and polaron pairs. By taking a semi-classical approach, we aimed to correct small scale *ab initio* simulations for being part of a larger system. In doing so, we found that for small molecule organic semiconductors, thermal motion likely has an insignificant effect on the Coulomb interaction of a polaron pair. We have also find that Coulomb integrals are more reliable than charge centers for calculating the Coulomb contribution to the polaron pair binding energy, especially if the hole and electron wave functions start to overlap. Proper surface termination is well known to be critical to predictable electronic properties of interfaces. Our work underscores this fact, as the electronic properties of unhydrogenated interfaces are significantly different both without the pentacene molecule and with it present. The chemical reaction of pentacene and silicon creates completely new electronic states. The new electronic states defy the reasonable application of our semi-classical models for polaron pair binding energy. However, with good surface termination, hybrid polaron pair binding energies calculated with the methods here should be reasonably accurate. If these static models of polaron pair binding prove sufficient in most cases, then a clear path exists toward creating better kinetic models.

**Acknowledgements.**

In addition to the simulation software packages described in the method section, several free and open source software packages were used. The author would like to acknowledge NumPy[44] (www.numpy.org), Matplotlib[45] (matplotlib.org), and POV-Ray[46] (povray.org) used for analysis, 2D plotting, and 3D rendering in this work.

**Appendix A.**

An excited electron and hole on a small molecule could be idealized as two co-centered Gaussian charge densities. The Coulomb binding energy for this electron-hole pair is derived below. The charge density of the electron and hole are represented respectively by the Gaussian distributions.

$$\rho_e(r) = \frac{q_e}{\sigma_e^3 \sqrt{2\pi}^3} e^{-\frac{1}{2}\left(\frac{r}{\sigma_e}\right)^2}$$

$$\rho_h(r) = \frac{q_h}{\sigma_h^3 \sqrt{2\pi}^3} e^{-\frac{1}{2}\left(\frac{r}{\sigma_h}\right)^2}$$

Where $q_e$, $q_h$ and $\sigma_e$, $\sigma_h$ are the charges and distribution widths of the electron and hole respectively. The electrostatic potential field due to the electron is:

$$\phi_e(r) = \frac{1}{4\pi\epsilon} \frac{q_e}{r} erf\left(\frac{r}{\sqrt{2}\sigma_e}\right)$$

Where $\epsilon$ is the dielectric permittivity. Integrating the charge density of the hole with the electrostatic potential yields the Coulomb binding energy, $U$.

$$U = \int \rho_h(r)\phi_e(r) d^3r$$

$$U = \int_0^\infty \rho_h(r)\phi_e(r) 4\pi r^2 dr$$

$$U = \frac{q_e q_h}{\varepsilon \sqrt{2\pi}^3} \frac{1}{\sqrt{\sigma_h^2 + \sigma_e^2}} = \frac{q_e q_h}{4\pi\varepsilon} \frac{1}{\sqrt{\frac{\pi}{2}(\sigma_h^2 + \sigma_e^2)}}$$

While this is only a simple approximation, it demonstrates how an electron and hole can have a finite binding energy with no observable distance between them.

**Appendix B.**

For a carrier near a dielectric interface, we define the interface as the x-y plane and positive z direction pointing into material 1. The solution for the electrostatic potential is adapted from Jackson's image charge analysis.[21] The electrostatic potential, $\phi^1$, of charge 1, $q_1$, is split into two parts: the potential in material 1, $\phi_1^1$, and the potential in material 2, $\phi_2^1$.

$$\phi^1(\vec{r}) = \begin{cases} \phi_1^1(\vec{r}), z > 0 \\ \phi_2^1(\vec{r}), z < 0 \end{cases}$$

$$\phi_1^1(\vec{r}) = \frac{q_1}{4\pi\epsilon_1}\left[\frac{1}{|\vec{r}-\vec{r}_1|} + \frac{\alpha_1}{|\vec{r}-\vec{r}_1\,'|}\right] \qquad \alpha_1 = \frac{\epsilon_1-\epsilon_2}{\epsilon_1+\epsilon_2}$$

$$\phi_2^1(\vec{r}) = \frac{q_1}{4\pi\epsilon_2}\left[\frac{\beta_1}{|\vec{r}-\vec{r}_1|}\right] \qquad \beta_1 = \frac{2\epsilon_2}{\epsilon_1+\epsilon_2}$$

The second term inside the brackets of the function $\phi_1^1(\vec{r})$ corresponds to the field from an image charge in material 2. The eletric field from charge 1 is also defined in the same piece-wise way as:

$$E_1^1(\vec{r}) = \frac{q_1}{4\pi\epsilon_1}\left[\frac{\vec{r}-\vec{r}_1}{|\vec{r}-\vec{r}_1|^3} + \alpha_1\frac{\vec{r}-\vec{r}_1\,'}{|\vec{r}-\vec{r}_1\,'|^3}\right]$$

$$E_2^1(\vec{r}) = \frac{\beta_1 q_1}{4\pi\epsilon_2}\left[\frac{\vec{r}-\vec{r}_1}{|\vec{r}-\vec{r}_1|^3}\right]$$

**Appendix C.**

A charge 1, $q_1$, in material 1 near the interface between material 1 and 2 induces a bound charge density at the interface between the two materials. This bound charge density Columbicly interacts with the original inducing charge 1. This self-polarization potential or self-energy, is the potential field associated with charge 1's image charge at $\vec{r}_1\,'$ acting on charge 1 at $\vec{r}_1$. As in Appendix B, the induced electric field in material 1 is:

$$E_1^{\sigma_1}(\vec{r}) = \frac{q_1\alpha_1}{4\pi\epsilon_1}\left[\frac{\vec{r}-\vec{r}_1\,'}{|\vec{r}-\vec{r}_1\,'|^3}\right]$$

Thus the electric field acting on charge 1 due to the bound interfacial charge density is:

$$E_1^{\sigma_1}(\vec{r} = \vec{r}_1\,') = \frac{q_1\alpha_1}{4\pi\epsilon_1}\left[\frac{\vec{r}_1-\vec{r}_1\,'}{|\vec{r}_1-\vec{r}_1\,'|^3}\right]$$

The distance vector, $\vec{r}_1 - \vec{r}_1\,'$, is mearly twice the distance to the interface.

$$\vec{r}_1 - \vec{r}_1\,' = 2h\hat{z}$$

$$E_1^{\sigma_1}(\vec{r}_1) = \frac{q_1\alpha_1\hat{z}}{16\pi\epsilon_1 h^2}$$

With this, we can calculate the change in the energy for charge 1 approaching the interface from infinitely far away to a distance $h_o$ to the interface and traveling infinity far into material 1 by integrating the force on charge 1. This is the self-polarization energy.

$$\Delta U_{Self} = \int_{\infty}^{h_o} -q_1 E_1^{\sigma_1} \cdot \hat{z} dh = -\frac{q_1^2 \alpha_1}{16\pi\epsilon_1 h_o}$$

Thus if $\alpha_1 > 0$, i.e. $\epsilon_1 > \epsilon_2$ then it is energetically favorible for charge 1 to leave the interface into material 1. Alternatively, if the opposite is true $\epsilon_1 < \epsilon_2$, then it will be favorable for charge 1 to move towards the interface. This result can be extended to diffuse classical charge density by means of coulumb integral of the charge desity with it's image density. However, in this context, the expecation value of this potential applied to a single particle wave function results in a simpler relation that is more in line with a quantum mechanical approach:

$$\Delta U_{Self} = -\frac{q_1^2 \alpha_1}{16\pi\epsilon_1} \int_{z>0}^{z\to\infty} \frac{|\psi_1(\vec{r}_1)|^2}{z} d^3\vec{r}_1$$

**Appendix D.**

The Coulomb interaction energy of two charges at on either side of the interface can found by the potential field of charge 1 acting on charge 2, $q_2$, at its position $\vec{r}_2$.

$$U_{charge-charge} = q_2 \phi_2^1(\vec{r}_2)$$
$$U_{charge-charge} = \frac{q_2}{4\pi\epsilon_2} \left[\frac{\beta_1 q_1}{|\vec{r}_2 - \vec{r}_1|}\right] = \frac{q_1 q_2}{4\pi(\epsilon_1+\epsilon_2)/2} \left[\frac{1}{|\vec{r}_2 - \vec{r}_1|}\right]$$

This result is interesting in its simplicity. As long as the charges are on opposite sides of the interface, it does not matter where; the interaction is Coulombic where the permittivity is average of the two materials. This result can also be easily extended to diffuse charges on either side of the interface:

$$U_{charge-charge} = \frac{1}{4\pi(\epsilon_1+\epsilon_2)/2} \int_{z>0}^{z\to\infty} d^3 r_1 \int_{z<0}^{z\to-\infty} d^3 r_2 \left[\frac{\rho_1(\vec{r}_1)\rho_2(\vec{r}_2)}{|\vec{r}_2 - \vec{r}_1|}\right]$$

The charge densities of the two diffuse charges are $\rho_1$ and $\rho_2$. This the same as the expectation value of the coulomb interaction between two uncorrelated, non-exchangeable particles.

## Appendix E.

We can approximate a polar molecule at the interface as a pair of opposite charges on either side with the same distance. This will start with the combined potential fields in material 1 and 2.

$$\vec{r}_2 = \vec{r}_1{}', \quad \vec{r}_1 = \vec{r}_2{}', \quad -q_1 = q_2$$
$$\phi_1(\vec{r}) = \frac{1}{4\pi\epsilon_1}\left[\frac{q_1}{|\vec{r}-\vec{r}_1|} + \frac{\alpha_1 q_1}{|\vec{r}-\vec{r}_2|}\right] + \frac{1}{4\pi\epsilon_1}\left[\frac{\beta_2 q_2}{|\vec{r}-\vec{r}_2|}\right]$$
$$\phi_2(\vec{r}) = \frac{1}{4\pi\epsilon_2}\left[\frac{\beta_1 q_1}{|\vec{r}-\vec{r}_1|}\right] + \frac{1}{4\pi\epsilon_2}\left[\frac{q_2}{|\vec{r}-\vec{r}_2|} + \frac{\alpha_2 q_2}{|\vec{r}-\vec{r}_1|}\right]$$

which simplfy to:

$$\phi_1(\vec{r}) = \frac{q_1}{4\pi\epsilon_1}\left[\frac{1}{|\vec{r}-\vec{r}_1|} + \frac{-1}{|\vec{r}-\vec{r}_2|}\right]$$
$$\phi_2(\vec{r}) = \frac{q_1}{4\pi\epsilon_2}\left[\frac{-1}{|\vec{r}-\vec{r}_2|} + \frac{1}{|\vec{r}-\vec{r}_1|}\right]$$

These can be treated with a dipole expansion:

$$\vec{d} = \vec{r}_1 - \vec{r}_2$$
$$\vec{p} = q_1 \vec{d}$$
$$\vec{r}_d = \frac{\vec{r}_1 + \vec{r}_2}{2}$$
$$\phi_1(\vec{r}) \approx \frac{1}{4\pi\epsilon_1}\frac{\vec{p}\cdot(\vec{r}-\vec{r}_d)}{|\vec{r}-\vec{r}_d|^3}$$
$$\phi_2(\vec{r}) \approx \frac{-1}{4\pi\epsilon_2}\frac{\vec{p}\cdot(\vec{r}-\vec{r}_d)}{|\vec{r}-\vec{r}_d|^3}$$

The molecular dipole moment is $\vec{p}$ and the location of the dipole is $\vec{r}_d$. If there is a thin sheet of these oreiented molecules at the interface due to texturing order, a net potential field is created. A diagram of this arrangement is avaible in Figure 14.

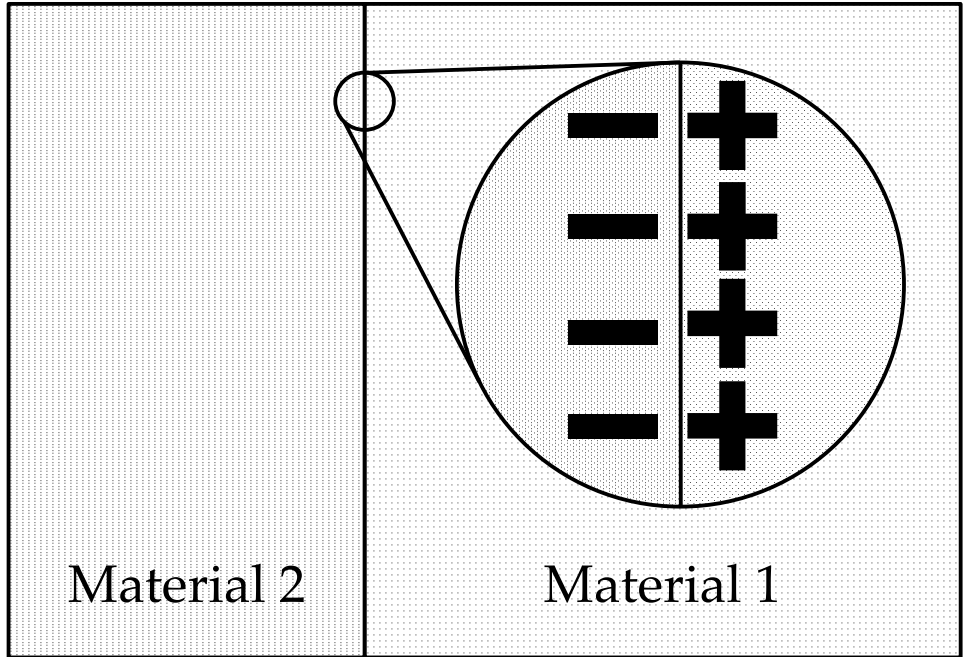

Figure 14. A thin layer of dipoles is arranged at the interface. They collectively act to form a potential field which is dependent on the dipole moment density.

The dipole density of this sheet in dipole moments $\vec{p}$, per area $A$, is given by $\vec{\sigma}$.

$$\vec{\sigma} = \frac{\vec{p}}{A}$$

The net potential field is given by integrating over interface.

$$\phi(\vec{r}) = \int \vec{g}(\vec{r} - \vec{r}_d) \cdot \vec{\sigma}(\vec{r}_d) \, d^2 r_d$$

where $\vec{r}_d$ is a position in the interface and $\vec{g}$ is the dipole field given by:

$$\vec{g}(\vec{r}) = \frac{1}{4\pi\epsilon} \frac{\vec{r}}{|\vec{r}|^3}$$

The permativity $\epsilon$ is $\epsilon_1$ in material 1 and is $\epsilon_2$ in material 2.

This integral can be switched to polar form for easier evaluation:

$$\rho = |\vec{r}_d|$$

$$\phi(\vec{r}) = \int_{\rho=0}^{\rho=\infty} \rho \, d\rho \int_{\theta=0}^{\theta=2\pi} d\theta \, \vec{g}(\vec{r} - \vec{r}_d) \cdot \vec{\sigma}(\vec{r}_d)$$

With a constant polarization density and substituting:

$$\phi(\vec{r}) = \frac{1}{4\pi\epsilon} \int_{\rho=0}^{\rho=\infty} \rho \, d\rho \int_{\theta=0}^{\theta=2\pi} d\theta \, \frac{1}{|\vec{r} - \vec{r}_d|^3} (\vec{r} \cdot \vec{\sigma} - \vec{r}_d \cdot \vec{\sigma})$$

There is no z component to $\vec{r}_d$ since it is just a sheet.

$$\vec{r}_d \cdot \vec{\sigma} = \rho |\vec{\sigma}| cos(\theta_\sigma - \theta)$$

where $\theta_\sigma$ is the angle of projection of the dipole moments onto the x-y plane.

$$\phi(\vec{r}) = \frac{1}{4\pi\epsilon} \int_{\rho=0}^{\rho=\infty} \rho d\rho \left( \int_{\theta=0}^{\theta=2\pi} d\theta \frac{\vec{r}\cdot\vec{\sigma}}{|\vec{r}-\vec{r}_d|^3} - \int_{\theta=0}^{\theta=2\pi} d\theta \frac{\rho|\vec{\sigma}|cos(\theta_\sigma - \theta)}{|\vec{r}-\vec{r}_d|^3} \right)$$

Since our system is translationally ivarient in the x and y directions, we only need consider the z component of our position vector:

$$\vec{r} = h\hat{z}$$
$$|h\hat{z} - \vec{r}_d| \to (h^2 + \rho^2)^{\frac{1}{2}}$$

and the potential becomes:

$$\phi(h) = \frac{1}{4\pi\epsilon} \int_{\rho=0}^{\rho=\infty} \rho d\rho \left( \int_{\theta=0}^{\theta=2\pi} d\theta \frac{h\hat{z}\cdot\vec{\sigma}}{(h^2+\rho^2)^{\frac{3}{2}}} - \int_{\theta=0}^{\theta=2\pi} d\theta \frac{\rho|\vec{\sigma}|cos(\theta_\sigma - \theta)}{(h^2+\rho^2)^{\frac{3}{2}}} \right)$$

the two interior integrals can be evaluted easily since the first has no dependce on $\theta$ and the second evalutes to zero. Also the only $z$ component of $\vec{\sigma}$ maters so we can replace it

$$\phi(h) = \frac{1}{4\pi\epsilon} \int_{\rho=0}^{\rho=\infty} \rho d\rho \ 2\pi \frac{h\hat{z}\cdot\vec{\sigma}}{(h^2+\rho^2)^{\frac{3}{2}}} = \frac{\sigma_z}{2\epsilon}\frac{|h|}{h}$$

thus:

$$\phi_1 = \frac{\sigma_z}{2\epsilon_1}$$
$$\phi_2 = -\frac{\sigma_z}{2\epsilon_2}$$

The jump across the interface is then:

$$\Delta\phi_{2\to1} = \frac{\sigma_z}{2}\left[\frac{1}{\epsilon_1} + \frac{1}{\epsilon_2}\right]$$
$$\Delta\phi_{1\to2} = -\frac{\sigma_z}{2}\left[\frac{1}{\epsilon_1} + \frac{1}{\epsilon_2}\right]$$

If there is pair of charges across this ordered interface, the enegry is then:

$$U_{inter} = \frac{\sigma_z}{2}\left[\frac{q_1}{\epsilon_1} - \frac{q_2}{\epsilon_2}\right]$$